\documentclass[5p]{elsarticle}

\usepackage[T1]{fontenc}
\usepackage{lmodern}
\DeclareFontFamily{OMX}{lmex}{}
\DeclareFontShape{OMX}{lmex}{m}{n}{<-> lmex10}{}

\usepackage{lineno,hyperref,subcaption,hyphenat}
\usepackage{caption}
\usepackage{amsmath, amsfonts, bm, mathtools}
\usepackage{bigints}
\usepackage{siunitx}
\usepackage{microtype}
\usepackage{wrapfig}
\usepackage{xspace}
\usepackage{placeins}
\usepackage{algorithm}
\usepackage[noend]{algpseudocode}

\modulolinenumbers[5]
\makeatletter
\def\BState{\State\hskip-\ALG@thistlm}
\makeatother
\usepackage[dvipsnames]{xcolor}
\newcommand\myshade{85}
\colorlet{mylinkcolor}{BrickRed}
\colorlet{mycitecolor}{NavyBlue}
\colorlet{myurlcolor}{Aquamarine}
\hypersetup{
  linkcolor  = mylinkcolor!\myshade!black,
  citecolor  = mycitecolor!\myshade!black,
  urlcolor   = myurlcolor!\myshade!black,
  colorlinks = true,
}

\newcommand\figsubref[2]{\hyperref[#1]{\ref*{#1}#2}}

\usepackage{ctable}

\begin{document}

\begin{frontmatter}

\title{Bayesian at heart: Towards autonomic outflow estimation \\ via generative state-space modelling of heart rate dynamics}

\author[Sussex,Psychedelics,Complexity,Eudaimonia]{Fernando E. Rosas}\fnmark[1]
\ead{f.rosas@sussex.ac.uk}
\author[PBI]{Diego Candia-Rivera}\fnmark[1]
\author[Cam,MNI]{Andrea I Luppi}\fnmark[1]
\author[hkust]{Yike Guo}\fnmark[1]
\author[Computing,CamPsy]{Pedro A.M. Mediano}\fnmark[1]

\address[Sussex]{School of Engineering and Informatics, University of Sussex, United Kingdom}
\address[Psychedelics]{Centre for Psychedelic Research, Department of Brain Science, Imperial College London, United Kingdom}
\address[Complexity]{Centre for Complexity Science, Imperial College London, London, United Kingdom}
\address[Eudaimonia]{Centre for Eudaimonia and Human Flourishing, University of Oxford, United Kingdom}
\address[PBI]{Paris Brain Institute, ICM, 75013, Paris, France}
\address[Cam]{University Division of Anaesthesia, University of Cambridge, Cambridge, United Kingdom}
\address[MNI]{Montreal Neurological Institute, McGill University, Montreal, Canada}
\address[hkust]{Department of Computer Science and Engineering, Hong Kong University of Science and Technology, Hong Kong}
\address[Computing]{Department of Computing, Imperial College London, South Kensington, London, United Kingdom}
\address[CamPsy]{Department of Psychology, University of Cambridge, Cambridge, United Kingdom}

\begin{abstract}

Recent research is revealing how cognitive processes are supported by a complex interplay between the brain and the rest of the body, which can be investigated by the analysis of physiological features such as breathing rhythms, heart rate, and skin conductance. 
Heart rate dynamics are of particular interest as they provide a way to track the sympathetic and parasympathetic outflow from the autonomic nervous system, which is known to play a key role in modulating attention, memory, decision-making, and emotional processing. 
However, extracting useful information from heartbeats about the autonomic outflow is still challenging due to the noisy estimates that result from standard signal-processing methods. To advance this state of affairs, we propose a paradigm shift in how we conceptualise and model heart rate: instead of being a mere summary of the observed inter-beat intervals, we introduce a modelling framework that views heart rate as a hidden stochastic process that drives the observed heartbeats. Moreover, by leveraging the rich literature of state-space modelling and Bayesian inference, our proposed framework delivers a description of heart rate dynamics that is not a point estimate but a posterior distribution of a generative model. 
We illustrate the capabilities of our method by showing that it recapitulates linear properties of conventional heart rate estimators, while exhibiting a better discriminative power for metrics of dynamical complexity compared across different physiological states.

\end{abstract}


\end{frontmatter}

\section{Introduction}

The autonomic nervous system (ANS) is responsible for the physiological adjustments necessary for the regulation of the body and contributes to appropriately responding to environmental demands. 
The ANS is composed of both afferent and efferent pathways that involve multiple neural structures including the vagus nerve, medulla, brainsteam, insula and other cortical and sub-cortical regions~\cite{berntson2019interoception}, and the result of their interactions constitutes what is known as \textit{autonomic outflow}~\cite{acharya_heart_2006, beissner_autonomic_2013}. 
Crucially, the impact of sympathetic and parasympathetic outflow goes far beyond driving homeostatic regulation and ``flight or fight'' responses, being also involved in cognitive processes such as attention, memory, decision-making, and emotional processing ~\cite{thayer_beyond_2006}. Thus,  measurement of the autonomic outflow can provide valuable insights into the physiological substrates that support human cognition~\cite{candia-rivera_brain-heart_2022}.

A popular approach to investigate autonomic outflow is by analysing properties of how heart rate changes over time. More specifically, a large body of literature has focused on the investigation of \textit{heart rate variability} (HRV)~\cite{clifford2002signal} --- i.e. the patterns of variation between the time intervals between successive heartbeats --- which is thought to reflect the changes in the sympathetic and parasympathetic branches of the autonomic nervous system~\cite{acharya_heart_2006}. The richness of this variability is not fully described merely by the standard deviation of the inter-beat intervals, and different markers have been proposed to capture a range of dynamic patterns indicative of health conditions and cognitive states~\cite{sassi_advances_2015}. While conventional metrics of HRV are based on spectral decomposition or other linear methods, these methods are unable to capture non-linear fluctuations in heart rate --- which have been shown to convey valuable information~\cite{acharya2004heart,shaffer2017overview}. 
Building on this principle, there is a progressively growing literature that provides complementary insights on HRV patterns by investigating fractality and  signal complexity via metrics such as  permutation entropy~\cite{cysarz2013quantifying,aziz2005multiscale} and detrended fluctuation analysis~\cite{penzel2003comparison,castiglioni2008local}, just to name a few. 
Overall, researchers have a rich set of tools to use cardiac signals to investigate autonomic outflow.

This promising state of affairs is, however, overshadowed by the fact that extracting useful information from cardiac data is highly non-trivial. 
Heart rate is typically conceived to be a summary statistic (i.e. a simplified description) of the inter-beat interval time series, the estimation of which is based on frequentist principles and results in a point estimate. 
This approach, however, has critical limitations: (i) it is not able to incorporate relevant prior knowledge (e.g. what type of heart rate values are physiologically plausible), (ii) it is unable to deliver metrics of confidence of the estimation, and (iii) its value relies only on the inter-beat interval of the present moment and hence estimation errors in inter-beat intervals result in important overestimations of heart rate variability. 
Unfortunately, a poor estimation of heart rate dynamics greatly hinders the capabilities of any downstream analysis. Overall, there is a great need for more powerful and flexible estimation methods that may allow us to better characterise the richness of heart rate dynamics~\cite{camm1996heart}.

To address these important limitations, in this paper we propose to rethink the process of estimating heart rate dynamics to better capture the autonomic outflow. 
For this purpose, we introduce a new framework to model and estimate heart rate dynamics based on state-space modelling~\cite{durbin2012time} and Bayesian statistics~\cite{gelman1995bayesian}. 
Using state-space modelling principles, our framework conceives the heart rate not as a summary statistic but as a hidden (i.e. not directly measurable) process which drives the observed sequence of heart beats. Moreover, using Bayesian techniques our approach delivers not a point estimate but a posterior distribution, which encodes the likelihood of possible heart rate trajectories given an observed sequence of inter-beat intervals. 
To illustrate the capabilities of the proposed framework, we use the posterior distribution to build Bayesian estimates of non-linear properties of heart rate dynamics. We show that these estimates exhibit higher discriminative power to distinguish between different physiological states. 
Our proposed framework expands researchers' toolkit to study the autonomic outflow, opening new opportunities to develop effective biomarkers for different conditions --- e.g. risk stratification of cardiovascular conditions or tracking the underlying physiology of embodied cognitive processing.

\section{Results}

\subsection{A Bayesian approach to model heart rate dynamics}

The conventional method to calculate heart rate involves inferring how many beats one would expect per minute on average given the observation of $N_\text{b}$ beats over a period of time of $T$ seconds, which leads to the estimate $\text{HR}=60 N_\text{b} / T$. 
If one is interested in a dynamical description of how the heart rate fluctuates over time, one can follow the same rationale and reduce the time period to the limit where $N_\text{b}\to1$ and $T$ becomes equal to the inter-beat interval $I_\text{b}$,  leading to the following estimate of the ``instantaneous'' heart rate:
\begin{equation}\label{eq:HRinst}
\text{HR}_\text{freq}(t) = \frac{60}{I_\text{b}(t)}.
\end{equation}
From a statistical perspective, this expression can be understood as the outcome of an elementary frequentist method of inference that delivers a point estimate for the average number of beats per minute --- in fact, it is the number of beats one would see if all beats were separated by the same inter-beat interval $I_\text{b}$. As such, it has the strengths and weaknesses of frequentist approaches: it is conceptually simple and computationally lightweight, although it cannot estimate its own uncertainty or incorporate prior knowledge on plausible heart rate values. Furthermore, as $\text{HR}_\text{freq}(t)$ ignores previous inter-beat interval values, errors in the estimation of $I_\text{b}(t)$ inevitably lead to overestimations of heart rate fluctuations.

\begin{figure}[ht]
  \centering
  \includegraphics{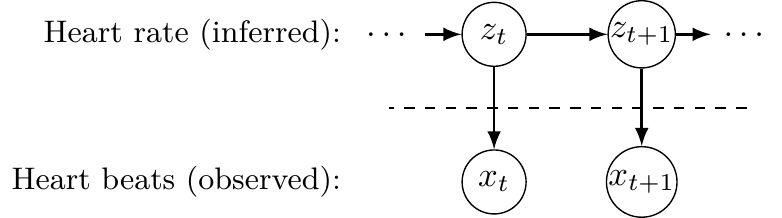}
  \caption{\textbf{Proposed heart rate state-space modelling approach}. The observable data (the heart beats) are assumed to be driven by the dynamics of a hidden stochastic process (the heart rate), which cannot be directly measured but can be inferred from the data.}
  \label{fig:SS}
\end{figure}

In contrast, our proposed approach conceives the heart rate as a hidden process that drives the actual observed heart beats, the statistical properties of which can be estimated via generative modelling.
Our model involves two time series corresponding to the
values of a dynamical process sampled with sampling frequency $f_\text{s} =
1/\Delta t$: 
$x_t$, which counts the number of heart beats that take place
during a temporal bin of length $\Delta t$, and $z_t$, which is the heart rate that drives the corresponding heart beats (Figure~\ref{fig:SS}). 
Our framework comprises a generative statistical model (see \textit{Methods}), the key component of which is a probability distribution $p$ that describes the likelihood of observing a given sequence of heart beats $x_1,\dots,x_N$ together with a heart rate time series $z_1,\dots,z_N$ . 

\begin{figure*}[ht]
\centering
  \includegraphics{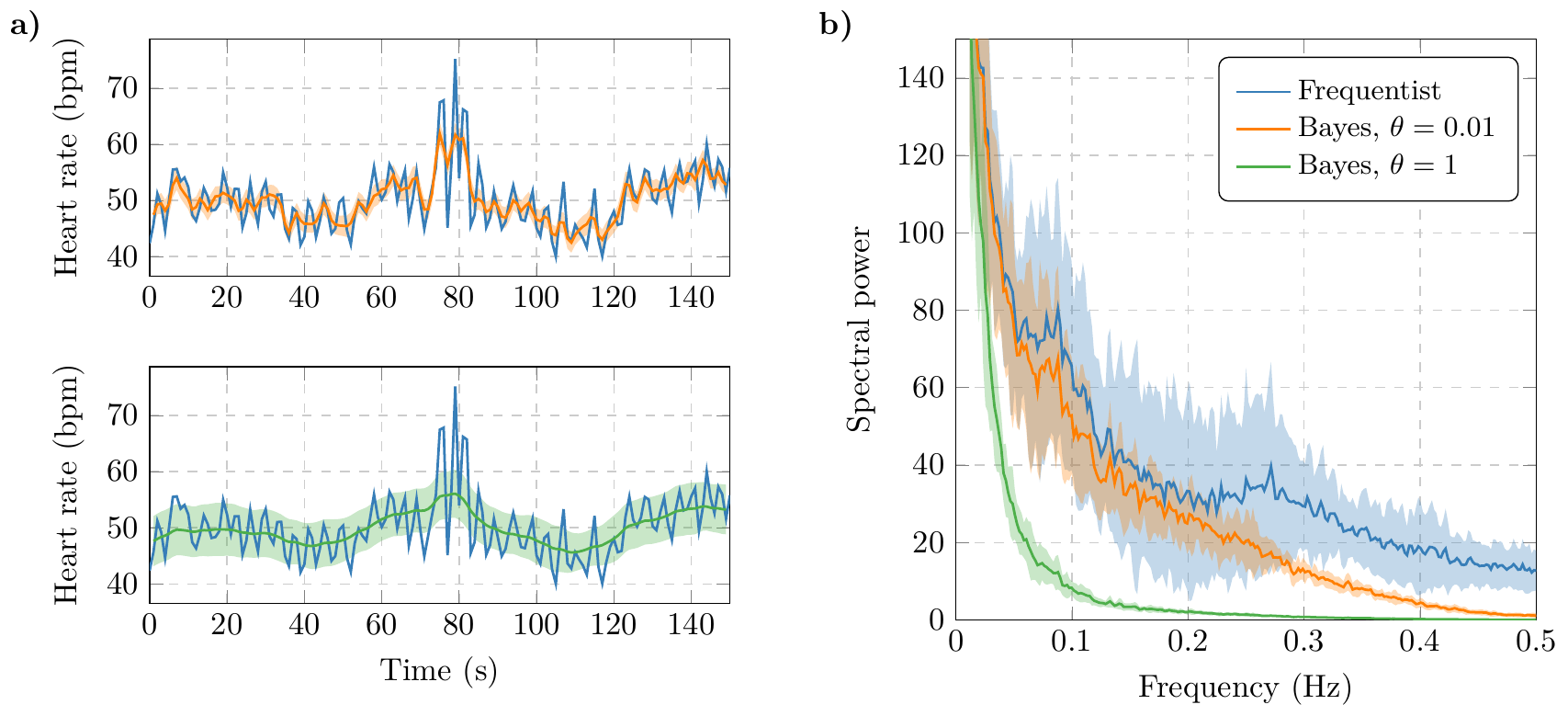}
\caption{\textbf{Illustration of the proposed method in example time series and power spectrum}. \textbf{a)} Comparison between frequentist (blue) and Bayesian (orange and green) estimation of heart rate, of which we display the mean and standard deviation obtained from 100 trajectories sampled from models with a weaker (top, $\theta=0.01$) and stronger (bottom, $\theta=1$) coupling strength between successive samples (see Section~\ref{sec:sampling}). \textbf{b)} Power spectrum of heart dynamics as obtained from frequentist and Bayesian approaches.}
  \label{fig:spectra}
\end{figure*}

Through this model, heart rate dynamics are now described not by a point estimate (i.e. as a single, most likely trajectory) but as obeying the following conditional distribution:
\begin{equation}
\text{HR}_\text{bayes}: \quad z_1,\dots,z_N \sim  p(z_1,\dots,z_N|x_1,\dots,x_N) .
\end{equation}
This posterior distribution describes the most likely heart rate trajectories $z_1,\dots,z_N$ given the observed data $x_1,\dots,x_N$. 
Crucially, by following methods proposed 
in Ref.~\cite{gugushvili2018fast} we avoid the computationally intensive task of computing the explicit posterior distribution, and instead use a Gibbs sampler~\cite{bishop2006pattern} to efficiently obtain sample trajectories. 
This allows not only to find the most likely trajectory, but also to estimate uncertainty (e.g. via the posterior variance). 
These sampled trajectories also allow us to build accurate estimators of non-linear properties of heart rate dynamics --- as developed in later sections.

\subsection{Method validation}

We tested our proposed method by analysing the autonomic outflow of healthy subjects going through a standard tilt-table protocol. This protocol places participants at various angles to monitor changes in cardiovascular activity
in order to assess the balance between the sympathetic and parasympathetic branches of the autonomic nervous system~\cite{porta_assessment_2007}. In particular, when the head is tilted upwards then blood flow to the brain decreases, which triggers activation of the sympathetic nervous system and suppression of the parasympathetic nervous system~\cite{cooke_human_1999,montano_power_1994}.

We used a public dataset\footnote{\href{https://physionet.org/content/prcp/1.0.0/}{https://physionet.org/content/prcp/1.0.0/}} comprising 10 healthy subjects (5 males, average age 28.7 $\pm$ 1.2 years) undergoing the tilt-table test~\cite{goldberger_physiobank_2000,heldt_circulatory_2003,heldt_computational_2002}.
Each subject was measured under four conditions (self stand up, slow tilt, fast tilt, and resting baseline) with a one-lead ECG. The resulting time series were used to calculate inter-beat intervals via template-based matching of QRS waves~\cite{candia-rivera_role_2021}. The sequence of inter-beat intervals was then used to construct two generative models for each subject, using two different values for a free parameter denoted by $\theta$ that regulates the variability between successive values of heart rate (see \textit{Methods}) --- which is closely related to the bandwidth of the resulting trajectories --- and sampled 100 heart rate trajectories from the posterior. Our approach worked as expected, generating smoother and more plausible trajectories
than the frequentist approach. The resulting trajectories and corresponding spectrum are shown in Figure~\ref{fig:spectra}.

\begin{figure}[h!]
\centering
  \includegraphics{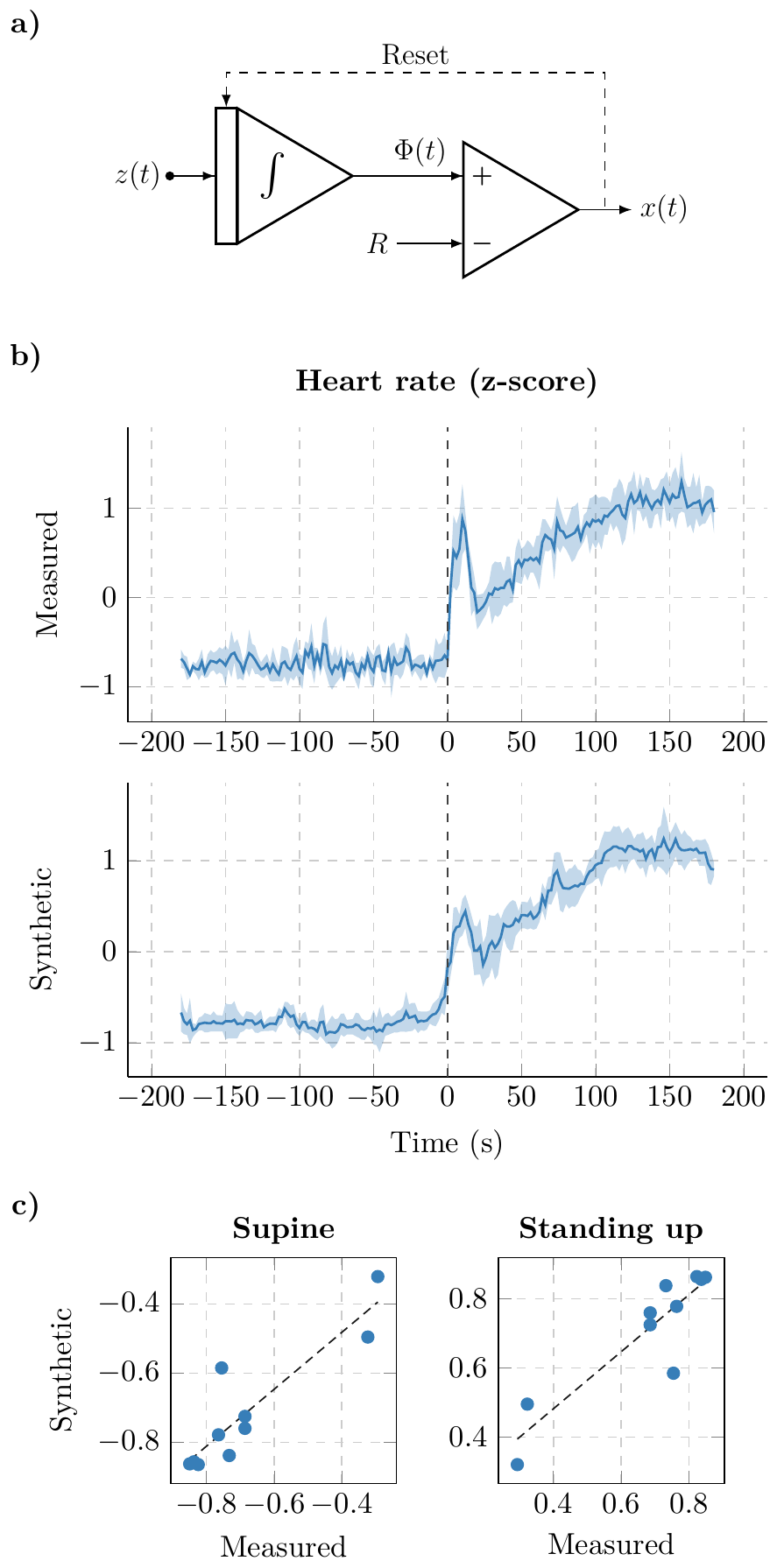}
  \caption{\textbf{Method validation}. \textbf{a)} Block diagram of the integral pulse frequency modulation (IPFM) model, which is used to generate synthetic heartbeats (see \textit{Methods}). \textbf{b)} Group-wise measured and synthetic heart rate series for the change from supine to upright position. The displayed signals correspond to the group median and shaded areas to the median absolute deviation. \textbf{c)} Scatter plot showing the correlation between measured and synthetic heart rate, in supine and standing up conditions. Dashed black lines are least squares regression lines. All the standing up postural changes were averaged per subject, centred at the time of postural change. Z-scores were computed per subject in the -180 to 180 s interval with respect to the postural change timing.}
  \label{fig:ipfm}
\end{figure}

To further test our proposed method, we used computational modelling to assess the consistency of the estimated autonomic outflow. For this purpose, we applied the average of the resulting trajectories as an input to a physiological model that emulates the autonomic stimulation to the sinoatrial node --- which is the main cardiac pacemaker~\cite{levy_neural_1981}. As an output, this model generates synthetic heartbeats via an Integral Pulse Frequency Modulation (IPFM) model (Figure~\figsubref{fig:ipfm}{a}) that takes the autonomic outflow estimation as an input for an integrate-and-fire process~\cite{candia2021_integral}, whose details are described in \textit{Methods}.

We used the obtained sequence of ``synthetic'' heartbeats to calculate heart rate using the simple frequentist approach given in Eq.~\eqref{eq:HRinst}, and studied its suitability by comparing it with the frequentist estimation of heart rate obtained from the original heartbeats (Figure~\figsubref{fig:ipfm}{b}). 
Results indicate that the synthetic heart rate matches the measured heart rate both when in supine position (z-score, measured: $1.12 \pm 0.11$; synthetic: $1.12 \pm 0.10$) and standing up (z-score, measured: $1.29 \pm 0.08$; synthetic: $1.30 \pm 0.08$). Moreover, the resulting values of mean heart rate were significantly correlated among subjects (Spearman $R = 0,8667$, $p = 0.0027$, Figure~\figsubref{fig:ipfm}{c}). Furthermore, the observed changes between conditions in the synthetic heart rate recapitulate the ones exhibited in the real data, as observed via a  paired Wilcoxon signed-rank test ($p = 0.0051$).

\subsection{Comparable discriminative power of spectral properties}

\begin{figure*}[ht]
  \centering
  \includegraphics{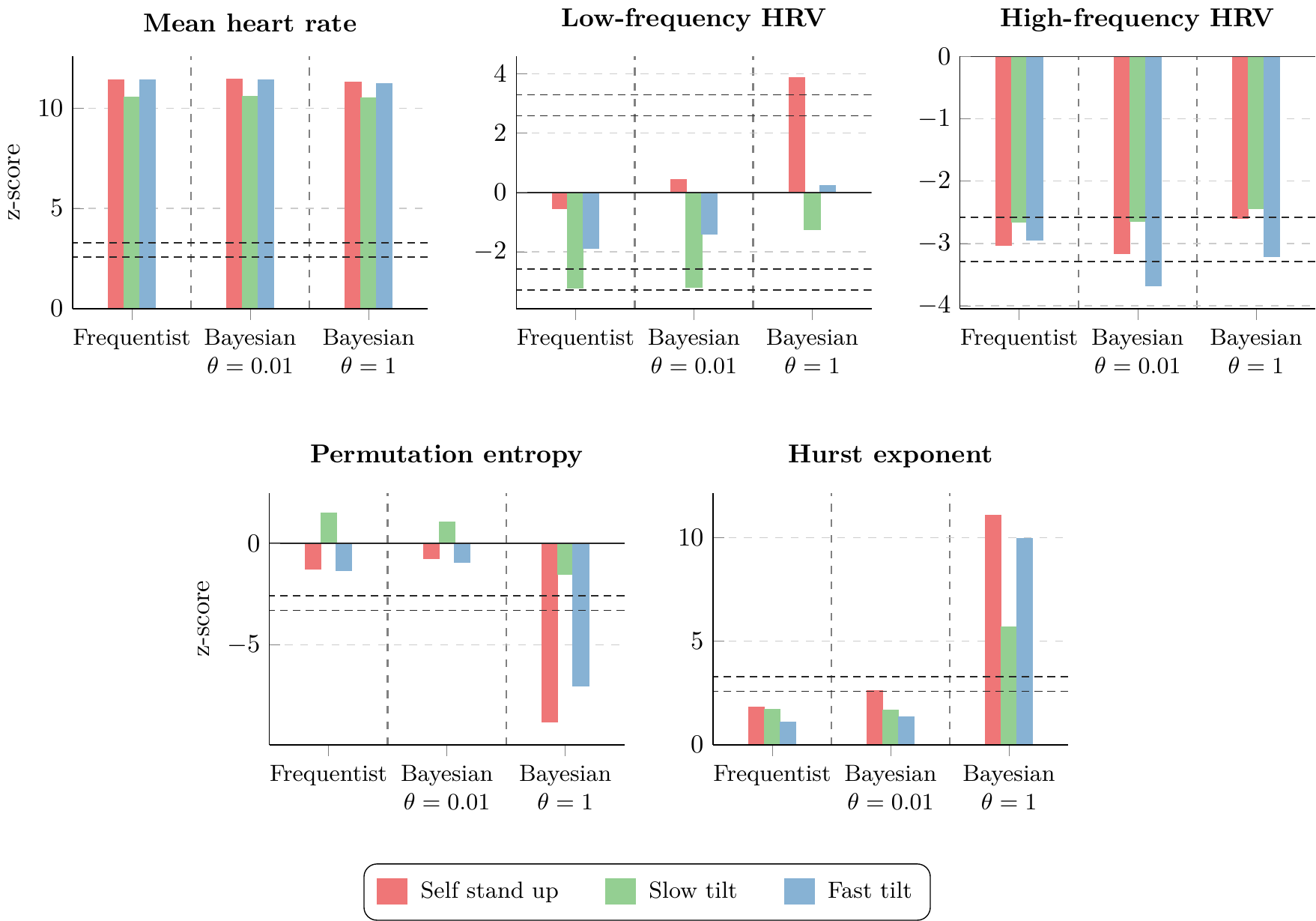}
  \caption{\textbf{Effects (in the form of z-scores) obtained from linear mixed-effect modelling of the effect of various physiological states on properties of heart rate dynamics}. Top row contains mean heart rate and spectral measures, and bottom row contains measures of dynamical complexity. While the Bayesian model with $\theta=0.01$ exhibits a similar discriminative power than the frequentist estimator for mean heart rate and spectral HRV, the Bayesian model with $\theta=1$ performs better for permutation entropy and Hurst exponent. Black horizontal lines represent p-value thresholds corresponding to $p = 0.01$ and $p = 0.001$.}
  \label{fig:lme}
\end{figure*}

The power of generative modelling can be harnessed to build Bayesian estimators of properties of heart rate dynamics. 
In particular, for a given property $F$ of a heart rate trajectory $z_1,\dots,z_N$ (e.g. entropy or spectral power), a Bayesian estimator can be built as follows:
\begin{equation}\label{eq:feature_posterior}
\hat{F} = \sum_{z_1,\dots,z_N} F(z_1,\dots,z_N) p(z_1,\dots,z_N|x_1,\dots,x_N)~,
\end{equation}
where the value of the property $F$ for each possible trajectory is weighted by the likelihood of such trajectory given the observed data.
Furthermore, if $F$ is a linear property, then Eq.~\eqref{eq:feature_posterior} accepts a shortcut: it reduces to $\hat{F} = F(\bar{z}_1,\dots,\bar{z}_N)$, where $\bar{z}_1,\dots,\bar{z}_N$ is the average trajectory under the posterior $p(z_1,\dots,z_N|x_1,\dots,x_N)$.

We sought to study spectral properties of heart rate dynamics via these Bayesian estimators, and investigated their capability to discriminate between the different physiological states in which the balance between sympathetic and parasympathetic activities is re-arranged. 
We compared the discriminative power of these Bayesian estimations against the obtained via the standard frequentist approach. 

For this purpose, we calculated Bayesian estimators for some linear features of heart rate dynamics, including the mean heart rate and spectral components of the heart rate variance, for each subject on each condition. To calculate spectral components, we first filtered the heart rate trajectories using a bandpass Butterworth filter of order 2 over standard HR bands (0.04--\SI{0.15}{\hertz} for low-frequency and 0.15--\SI{4}{\hertz} for high-frequency components), and then calculated the variance of the resulting signal. We refer to the variance of the low and high frequency signals as ``low-frequency HRV'' and ``high-frequency HRV,'' respectively.

We analysed the obtained values of mean heart rate, low-frequency HRV, and high-frequency HRV obtained for 
each stage and each subject via linear mixed-effect modelling. Specifically, we constructed models using the heart rate feature as target variable, stage as fixed effect, and modelled the effect of subject identity using a random intercept. 
We compared the effects observed with our Bayesian procedure using two different values for hyperparameter $\theta$: 0.01 (more loose) and 1 (more constrained). We contrasted these results with the ones obtained from a standard frequentist estimation. Results are shown in Figure~\ref{fig:spectra} and the resulting statistics are presented in \ref{app:stats}.

Overall, the effects observed for mean heart rate across conditions are very similar for all estimation approaches, while for HRV the model with $\theta=0.01$ presents small reductions and the one with $\theta=1$ presents stronger reductions --- consistent with the fact that $\theta$ is related to the bandwidth of the model (see \textit{Methods} and also Figure~\ref{fig:spectra}). Correspondingly, the model with $\theta=0.01$ achieves much better performance distinguishing between different states than the model with $\theta=1$, particularly  with high-frequency HRV. 
Moreover, it is worth noticing that the discriminative power of the model with $\theta=0.01$ is comparable --- and sometimes better --- than the one from the frequentist method.

\subsection{Better performance for measures of dynamical complexity}

It is generally recognised that analytic techniques inspired by principles from complexity science~\cite{waldrop1993complexity} are able to describe biological processes --- such as the ones that drive heart dynamics --- in illuminating ways~\cite{acharya_heart_2006}. 
When employing such techniques, it is worth noting that if a dynamical property of heart rate dynamics $F$ is non-linear, then its expected value $\hat{F}$ is in general different from the property evaluated on the average trajectory, $F(\bar{z}_1,\dots,\bar{z}_N)$.\footnote{For example, the well-known Jensen inequality states that, if $F$ is convex, then the mean value of $F$ is usually greater than $F$ evaluated in the mean value.} The fact that Bayesian estimators can assess this difference is one of their most powerful features. 
To illustrate how the proposed framework can be used to estimate metrics of dynamical complexity, in the following we consider two of them: the \textit{Hurst exponent} and \textit{sample entropy}. 

First, we consider the Hurst exponent of the heart rate time series as estimated via detrended fluctuation analysis (DFA)~\cite{peng1994mosaic} --- a method to determine the statistical self-affinity of a signal which is particularly useful for analysing time series with long-memory processes or $1/f$ noise~\cite{kantelhardt2001detecting} (see Ref.~\cite{ihlen2012introduction} for a tutorial). 
While DFA can be used just as a pragmatic way to obtain useful biomarkers, some researchers interpret its effectiveness as indicative of a fractal nature in heart rate fluctuations~\cite{eke2002fractal}.\footnote{Some argue that this would be related to the fractal structure of blood vessels and of nervous and humoral pathways connecting and hierarchically regulating local blood flows among several vascular beds~\cite{castiglioni2018multifractal}. However, the precise physiological bases of DFA are a matter of debate, and efforts to explain it have been based on relationships with spectral properties~\cite{francis2002physiological,willson2002relationship} and via computational modelling~\cite{rojo2007analysis}.} 
As a simple proof of concept, here we calculated the Hurst exponent via a simple application of DFA using the open-source package \texttt{Fathon}~\cite{bianchi2020fathon}.

As a second complexity measure, we consider the well-known permutation entropy~\cite{bandt2002permutation}, which analyses the patterns in the heart rate dynamics by classifying them into discrete classes and then studying their frequencies ---
being an example of a more general approach known as ``symbolic dynamics''~\cite{cysarz2013quantifying}.
Permutation entropy is a popular technique for measuring the pattern complexity in time series analysis in general~\cite{zanin2012permutation} and heart rate dynamics in particular~\cite{bian2012modified}, with extensions that can assess multiple temporal scales~\cite{aziz2005multiscale}. For the sake of simplicity, we performed an elementary calculation of permutation entropy using the open source package \texttt{Ordpy}~\cite{pessa2021ordpy}.

Both Hurst exponent and permutation entropy were calculated on each of the 100 sampled trajectories per subject, and then averaged to provide the outcome of the proposed Bayesian estimator. For comparison, we also evaluated both properties in the point estimate trajectory provided by the frequentist approach. 
Results show that neither the frequentist method or the Bayesian model with $\theta=0.01$ are able to find significant differences in either metric across conditions. Crucially, however, the Bayesian model with $\theta=1$ does reveal strong effects in both metrics, particularly in the self stand up and fast tilt conditions, showing that our approach yields more suitable heart rate time series for downstream complexity analyses.
%

\section{Discussion}

This paper introduces a new conceptual and modelling paradigm for the assessment of autonomic processes, in which the heart rate is construed as a hidden stochastic process that drives the generation of observed heart beats. Our framework instantiates this idea using the rich toolkit of state-space modelling and Bayesian statistics, with efficient algorithms to sample the resulting model and generate corresponding heart rate trajectories.

We showed how sampled trajectories can be used to estimate the posterior distribution of various properties of heart rate dynamics, leading to Bayesian estimators of spectral power or dynamical complexity. In agreement with the literature~\cite{porta_assessment_2007}, our results on the autonomic effects in a tilt-table test revealed a transient increase in heart rate and decrease in HRV in response to standing upright that is observed in both Bayesian and standard (frequentist) estimates of autonomic outflow. 
Importantly, results further show that the proposed Bayesian estimators of dynamical complexity measures such as permutation entropy and the Hurst exponent exhibit more discrimative power than the corresponding estimates obtained from the frequentist approach.

Our approach for estimating autonomic outflow ought to be seen as a first instantiation of a powerful and flexible paradigm, which can be significantly extended in the future. 
For example, the proposed model consists of a hidden process composed of a single time series, which aims to capture the aggregated autonomic outflow. Future work could use this framework to model sympathetic and parasympathetic activities separately, as well as to model the interactions between multiple physiological signals. This could help identify the neural substrates of underlying brain-heart interactions~\cite{candia-rivera_functional_2022, candia-rivera_modeling_2023}, and capture the specific nodes from the central autonomic network that are involved~\cite{beissner_autonomic_2013}.
Another limitation of the presented approach is that it does not account for the separate impact of blood pressure and respiration on heart rate~\cite{elghozi_effects_1991}, which could introduce biases in the results.

It is worth noting that our modelling approach focuses on modelling heart rate dynamics and not heart beats, which are taken as input data for estimating the former. 
Other approaches are more appropriate for modelling and predicting inter-beat intervals or heartbeat events, such as the methods developed in Refs.~\cite{barbieri_point-process_2005,valenza2018measures,sherman2022cardiac}. 
In contrast, our approach focuses on the posterior distribution of the heart rate dynamics given the observed heart beats.

Another relevant feature of the proposed modelling approach is that it is well-suited to the analysis of non-stationary autonomic data. 
Heart beat dynamics can be highly non-stationary in certain conditions, e.g. under postural changes (as in the tilt-table protocol considered in our analyses). 
Accordingly, our proposed method builds a model that is non-stationary by construction, and is hence fully capable of accounting for non-stationary features of heart rate dynamics. The exploitation of this powerful feature is the subject of ongoing work.

Overall, the present paper puts forward a new conceptual and practical approach to model autonomic dynamics which offers several new research opportunities and potential applications. 
While here we present simple validation and proof-of-concept, further work 
is required to delineate the precise physiological implications of the proposed construction.  
The proposed framework may lead to practically useful biomarkers to diagnose and monitor clinical conditions, while providing a basis for future modelling and simulation studies of the autonomic system and the involved neuronal mechanisms underlying physiological regulation.

\section{Materials and Methods}


\subsection{Overall modelling approach}

Our model of heart rate dynamics is based on two time series that correspond to the
values of dynamical processes with sampling frequency $f_\text{s} =
1/\Delta t$: 
$x_t$, which counts the number of heart beats in a temporal bin of length $\Delta t$ (i.e. between the
present moment and the previous sample), 
and $z_t$, which represents the heart rate that drives the corresponding heart beats (Figure~\ref{fig:SS}). 
In general, we assume that the heart beats $x_t$ are observed, and that sequence of heart rates $z_t$ follows a hidden (i.e. unmeasured) stochastic process whose statistical properties can nevertheless be inferred. 

In order to build a joint probability distribution over heart beats $x_1,\dots,x_N$ and heart rates $z_1,\dots,z_N$, 
we follow the state-space literature~\cite{durbin2012time} in adopting the following assumptions that make inference of the hidden process tractable:
\begin{itemize}
\item[(i)] the dynamics of the hidden process are Markovian; and
\item[(ii)] given the value of $z_t$, the observable $x_t$ is conditionally independent of all other observations and values of the hidden process.
\end{itemize}
Thanks to these assumptions, the joint distribution of heart rate and heart beat counts can be expressed as:
\begin{equation}
p(x_1,z_1,\dots,x_T,z_T) = p(z_1)p(x_1|z_1) \prod_{t=2}^T p(z_t|z_{t-1}) p(x_t|z_t)
\end{equation}
Therefore, the specification of the full model requires only three ingredients: the heart rate dynamics in the form of the conditional probability $p(z_t|z_{t-1})$, the link between heart rate and heart beats $p(x_t|z_t)$, and the distribution of the initial condition $p(z_1)$.

\subsubsection{Heart rate dynamics via Gamma Markov chains}
\label{sec:gmc}

Here we specify a generative model for heart rate dynamics, which follows the approach introduced in Ref.~\cite{gugushvili2018fast} for general point processes.

\begin{figure}[ht]
  \centering
  \includegraphics{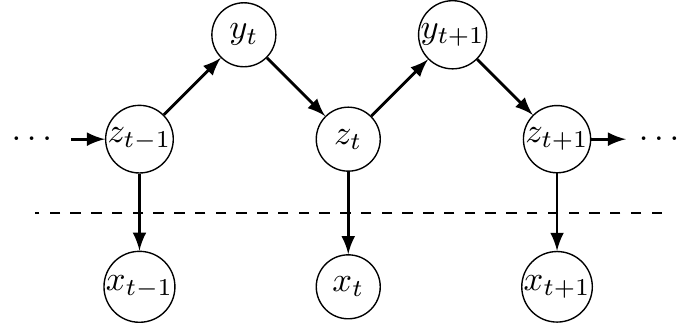}
  \caption{\textbf{Graphical model representation of the proposed state-space model for heart rate dynamics}. The actual heart rate is tracked by $z_1,\dots,z_T$, while the variables $x_1,\dots,x_T$ correspond to the count of heart beats over a given time interval, and $y_1,\dots,y_T$ are auxiliary variables.}
  \label{fig:SS_extended}
\end{figure}

Let us first specify the dynamics of the hidden process that models the heart rate, $p(z_t|z_{t-1})$ and $p(z_1)$. The most common instantiation of state-space models considers Gaussian
variables~\cite{durbin2012time}, however we avoid this approach as we heart rate is a non-negative quantity. Instead, our model considers the hidden dynamics as driven by a
Gamma Markov chain (GMC)~\cite{cemgil2007conjugate,dikmen2009unsupervised}, which builds the dynamics of the hidden process $z_1,\dots,z_T$ with the addition of auxiliary
variables $y_2,\dots,y_T$ whose joint distribution is a Markov chain of the
form (Figure~\ref{fig:SS_extended}):
\begin{equation}
  p(z_1,y_2,\dots,z_T,y_T) = p(z_1)\prod_{j=2}^T p(y_j|z_{j-1}) p(z_j|y_j),
\end{equation}
with the corresponding distributions are defined as follows:
\begin{align}
\begin{split}
z_1 &\sim \text{G}(z_1;\alpha_1,\beta_1),\\
y_t| z_{t-1} &\sim \text{IG}(y_t;\gamma,\gamma z_{t-1}),\\
z_t|y_t &\sim \text{G}\left(z_t;\gamma,\frac{\gamma}{y_{t}}\right),
\end{split}
\end{align}
where $\alpha_1,\beta_1$ and $\gamma$ are hyperparameters, and $\text{G}$ and $\text{IG}$ denote the Gamma and Inverse Gamma distributions, which are given by
\begin{align}
\begin{split}
\text{G}(z;\alpha',\beta') 
&=
\frac{\beta'^{\alpha'}}{\Gamma(\alpha')} z^{\alpha'-1} e^{-\beta' z},
\\ 
\text{IG}(y;\tilde{\alpha},\tilde{\beta}) 
&=
\frac{\tilde{\beta}^{\tilde{\alpha}}}{\Gamma(\tilde{\alpha})} 
z^{-\tilde{\alpha}-1} e^{-\tilde{\beta}/ z},
\end{split}
\end{align}
where $\alpha',\tilde{\alpha}$ and are shape parameters and $\beta',\tilde{\beta}$ are scale parameters. 
Crucially, this choice of auxiliary variables makes the full conditionals to have a simple form:
\begin{align}
\begin{split}
y_t|z_{t-1},z_t &\sim \text{IG}\left(y_t;2\gamma, \gamma x_{t-1} + \gamma x_t \right),
\\
z_t|y_{t},y_{t+1} &\sim \text{G}\left(z_t;2\gamma, \frac{\gamma}{y_{t}} + \frac{\gamma}{y_{t+1}} \right),
\\
z_1|y_{2} &\sim \text{G}\left(z_1;\alpha + \gamma, \beta + \frac{\gamma}{y_{t+1}} \right),
\\
z_T|y_{T} &\sim \text{G}\left(z_T;\gamma, \frac{\gamma}{y_{T}} \right),
\end{split}
\label{eq:posteriors1}
\end{align}
which enables an efficient sampling of heart rate trajectories, as explained in the next section.

Let us now specify the distribution that links heart rate and heart beats, $p(x_t|z_t)$. To leverage the powerful methods of Bayesian statistics, our natural choice is to consider a Poisson distribution --- the conjugate likelihood of the Gamma distribution. Hence,
\begin{equation}\label{eq:poisson}
x_t|z_t \sim P(x_t;z_t),
\end{equation}
where $P(x;\lambda) = \lambda^{x} e^\lambda/ x!$ is the Poisson distribution for rate $\lambda$. Please note that while the Poisson distribution generally yields a suboptimal fit to empirical distributions of the number of heart beats per bin in general,\footnote{For example, one can check that while the number of heart beats in a given time window grows linearly with bin length, the variance doesn't.} our usage of Poisson distributions avoids this issue --- as explained in the next section.

Finally, let's discuss the parameters of the model.
From Eqs.\eqref{eq:posteriors1} and \eqref{eq:poisson}, it can be seen that the model parameters are only $\alpha_1$ and $\beta_1$
for specifying $p(z_1)$, and $\gamma$ for determining the connectivity strenght
between sucessive samples of the hidden process. Our approach
is to choose $\alpha_1$ and $\beta_1$ by performing a
maximum likelihood estimation of the parameters of a Gamma distribution 
that would most likely have generated heart rate estimates as calculated by the
simple method given in Eq.~\eqref{eq:HRinst}.
For $\gamma$, we follow Ref.~\cite{gugushvili2018fast} and consider a prior distribution over it --- in this work we use $\gamma\sim \text{Exp}(\gamma;\theta)$ with $\text{Exp}$ denoting an exponential distribution, and $\theta$ is the hyperparameter of this prior.

\subsubsection{Sampling the generative model}
\label{sec:sampling}

A powerful feature of Bayesian statistics is that they allows efficient methods to sample the posterior distribution $p(z_1,\dots,z_T|x_1,\dots,x_T)$.  
This can be done via a \textit{Gibbs sampler}~\cite{geman1984stochastic,gelfand1990sampling}, a Markov chain Monte Carlo (MCMC)~\cite{brooks2011handbook} algorithm that can be used to obtain sequences of observations that follow a given probability distribution when direct sampling is difficult.
The Gibbs sampler provides an efficient procedure to extract trajectories from the posterior distribution, as we explain in the sequel.

Following Ref.~\cite{gugushvili2018fast}, one can show that the posterior distributions have --- thanks to the modelling choices --- the following simple expressions:
\begin{align}
\begin{split}
y_t|z_{t-1},z_t &\sim \text{IG}\left(y_t; 2\gamma, \gamma x_{t-1} + \gamma x_t \right),
\\
z_t|y_{t},y_{t+1},x_t &\sim \text{G}\left(z_t; 2\gamma + x_t, \frac{\gamma}{y_{t}} + \frac{\gamma}{y_{t+1}} + n\Delta t \right),
\\
z_1|y_{2},x_1 &\sim \text{G}\left(z_1; \alpha + \gamma + x_1, \beta + \frac{\gamma}{y_{t+1}} + n\Delta t \right),
\\
z_T|y_{T},x_T &\sim \text{G}\left(z_T; \gamma + x_T, \frac{\gamma}{y_{T}} + n\Delta t \right).
\end{split}
\label{eq:posterior}
\end{align}
Furthermore, the (unnormalised) posterior distribution for parameter $\gamma$
given $z_t,y_t$ can be shown to be the following non-standard distribution~\cite{gugushvili2018fast}:
\begin{align}
  \gamma | z_t,y_t
  \sim \; &
  \text{Exp}(\gamma;\theta) \times \left( \frac{\gamma^\gamma}{\Gamma(\gamma)}
  \right)^{2(T-1)} \prod_{j=2}^T \left( \frac{z_{j}z_{j-1}}{y_j^2}
  \right)^\gamma \nonumber
  \\
  & \times \exp\left( -\gamma \sum_{j=2}^T \frac{z_j+z_{j-1}}{y_j} 
  \right). \label{eq:fea}
\end{align}

Using the above equations, samples of trajectories can be computed as follows. 
First, the sampler is initialised using independent samples for $z_1,\dots,z_T$
drawn from a Gamma distribution with parameters that best
match a simple frequentist estimation of the heart rate (i.e. using the same values of $\alpha$ and $\beta$
obtained for $p(z_1)$, as specified in the previous section).
Also, $\gamma$ is initalised by the mean value of its prior, i.e. the value of
it hyperparameter $\theta$.
Using the obtained initial values of $z_t$ and $\gamma$, the sampler then iterates as follows: 
\begin{algorithm}
\begin{algorithmic}[1]
  \State \texttt{Estimate $y_t$ given $x_t$ and $\gamma$,
  \State Estimate $z_t$ given $y_t$ and $\gamma$,
  \State Estimate $\gamma$ given $z_t,y_t$.
  \State Go to $1$}
\end{algorithmic}
\end{algorithm}

In this procedure, the estimation of $z_t$ and $y_t$ is done directly using Eqs.~\eqref{eq:posterior}. 
For the estimation of $\gamma$ we follow Ref.~\cite{gugushvili2018fast} in using 
a Metropolis-Hastings MCMC using a zero-mean Gaussian with standard deviation $\tau$ as the proposal function. The value of $\tau$ was selected to achieve an intermediate number of accepted and rejected steps in the Markov chain, and results were empirically insensitive to many choices of $\tau$.
Also, please note that when sampling heart rate trajectories, there is no need to keep record of the values of the auxiliary variables $y_1,\dots,y_T$.

\subsubsection{Choosing parameter values}

The sampling procedure described in previous subsection has only two hyperparamenters, $\theta$ and
$\tau$, which are related with the prior and estimation of the connectivity strength $\gamma$. 
In our experiments samples were seen to be fairly insensitive
to $\tau$, while their smoothness strongly depended on $\theta$. For this reason, we set $\tau=1$ and consider two values of $\theta$: 0.01 for allowing more variation between successive heart rate values, and 1 to make their connectivity stronger (Figure~\ref{fig:SS}).

Another parameter to be chosen is the sampling frequency $f_s$. To determine this value, we followed two criteria: we would like to have a relatively
high sampling frequency to be able to assess heart dynamics with good
temporal resolution, and we would like to make the Poisson statistics of
$p(x_t|z_t)$ fit the distribution of empirical data as well as possible. 
Luckily, both criteria can be satisfied by adopting a relatively high frequency; in
this paper we use $f_\text{s}=\SI{3}{\hertz}$.
Indeed, the resulting small temporal bins means $x_t$ is effectively a binary variable, such that $x_t=1$ if
there was a heart beat during that bin and $x_t=0$ otherwise.
Furthermore, as the frequencies of events $x_t=1$ is relatively low
(approximately $1/3$ for a heart rate near 60 BPM), such
statistics can be well fit by a Poisson distribution with low rate.

This choice, however, has two downsides --- which luckily do not matter for the
current application. First, the relatively regular appearance of one beat every three bins
generates a small artefactual fluctuation. We deal with this in an heuristic
manner: we do a smoothing step (a rolling mean with a triangle window of 9
samples), and then a subsequent subsampling step keeping one every three
samples. With this, our resulting sampling frequency is $f_\text{s}=\SI{1}{\hertz}$, which is still high enough for the purpose of studying heart rate dynamics.  
The second downside is that the model cannot be used to sample meaningful
sequences of heart beats, as the conditional independence of the corresponding
variables is violated due to the small bin size. However, our goal is not to
sample heart beats but heart rate trajectories, and hence this limitation is not
important. Importantly, violations to this statistical condition do not imply that the modelling becomes inaccurate --- in fact, the efficacy of the posterior of the proposed model beyond this condition is equivalent to the well-known broad efficacy of the Naive Bayes classifier in a wide range of scenarios that don't satisfy its statistical assumptions~\cite{domingos1997optimality}.

\subsection{Building Bayesian estimators of properties of heart dynamics}

Finally, we can leverage this trajectory sampling technology to build Bayesian
estimators of properties of heart rate dynamics, as follows.

Let's use the shorthand notation $\bm z = (z_1,\dots,z_T)$ for a sampled trajectory of the heart rate
dynamics, and consider $F(\bm z)$ to be a scalar function of this trajectory.
For example, this could be an estimator of the trajectory's entropy or of its Hurtz exponent. Then, using our Gibbs sampler we could extract samples of the
posterior distribution $p(F(\bm z) | x_1,\dots,x_T)$. One can then use these
samples to estimate various useful features of this distribution, such as its mean or standard
deviation. 
In particular, the presented analyses estimate various properties of heart rate dynamics as the empirical mean value of this posterior.

Importantly, our experiments suggest that the proposed model for heart rate dynamics is generally non-ergodic. 
Therefore, our method to sample the posterior is as follows.
To generate one trajectory, we run the Gibbs sampler
$N_\text{r}$ iterations and discard the first $N_\text{d}$.
We calculate the mean value of the remaining $N_\text{r}-N_\text{d}$ runs and take the result as a single trajectory.\footnote{Technically, we are sampling the posterior distribution of the \textit{mean} heart rate trajectory. Given the non-ergodicity of the sampler, this approach was more empirically powerful than analysing non-averaged sampled trajectories directly.}
To generate $N_\text{s}$ trajectories, we run this procedure $N_\text{s}$ times, 
initialising the Gibbs sampler every time with new random initial conditions. 
For the presented results, for each time series of heartbeats we run the Gibbs sampler for $N_\text{r}=2\times 10^4$, and
discard the first $N_\text{d} = 5\times 10^3$.

\subsection{ECG data and preprocessing}

To validate our framework, we analysed the estimation of heart rate on a data set of postural changes. The data set comprises one-lead ECG series from 10 healthy subjects (5 males and 5 females, average age $28.7 \pm 1.2$ years) while undergoing the tilt-table test~\cite{goldberger_physiobank_2000,heldt_circulatory_2003,heldt_computational_2002}. The subjects were initially asked to remain in a horizontal supine position and to move to a vertical position with the help of either the tilt-table or by self-stand up. The subjects were part of six sessions that were sorted randomly between resting periods: two stand up, two slow tilts (\SI{50}{\second} from 0 to 70°), and two fast tilts (\SI{2}{\second} from 0 to 70°), while remaining in each condition for approximately \SI{3}{\minute}. The entire protocol lasted between 55 and 75 minutes. 

For simplicity of the analysis, we aligned all trials by subject between -180 to 180 seconds with respect to the change from supine position to standing up. 
ECG preprocessing involved frequency filtering, R-peak detection, and correction of misdetections. ECG data were bandpass-filtered using a Butterworth filter of order 4, between 0.5 and \SI{45}{\hertz}.
Heartbeats from QRS waves were identified in an automated process based on a template-based method for detecting R-peaks~\cite{candia-rivera_role_2021}.

\subsection{Validation of generated heart rate via an Integral Pulse Frequency Modulation (IPFM) model}

Computational modelling has been employed to generate synthetic series of heartbeat dynamics, which help us understand autonomic dynamics in different conditions~\cite{mcsharry_dynamical_2003}. Among the different models that have been proposed, the integral pulse frequency modulation (IPFM) models describe the physiological transduction from the autonomic outflow to heartbeat generation~\cite{mohn_modelling_1978}. These models are developed under the hypothesis of the existence of a heartbeat function as a real-time modulation function representing stimulation of the sinoatrial node, which is directly related to heartbeat generation~\cite{levy_neural_1981}. The modulation function in IPFM models considers the combination of heart rate components of sympathetic-vagal control as the inputs to an integrator that generates the heartbeats~\cite{bailon_integral_2011, candia2021_integral, brennan_poincare_2002}.

Mathematically, the heartbeat generation model is based on the temporal integration of the autonomic outflow, denoted by $z(t)$,\footnote{Note that we have changed notation from discrete-time $z_t$ in Sec.~\ref{sec:gmc} to continuous-time $z(t)$.} which contains the modulations from sympathetic and parasympathetic nervous systems. Given the previous beat occurred at time $t_k$, the model considers the following integral function:
\begin{equation}\label{integral}
  \Phi_k(t) = \int_{t_k}^{t} z(t) dt ~ .
\end{equation}
When $\Phi_k$ reaches a threshold R (which for simplicity we fix at $R=1$), 
then the $k+1$-st heartbeat is generated and the integral is reset. Finally, the heartbeat function $x(t)$ is modelled as a sum of Dirac functions $\delta(t)$ positioned at each heart beat timing $t_k$:
\begin{align}\label{dirac}
  x(t) = \sum_{k} \delta(t - t_k) ~ .
\end{align}

Note that, in principle, one could formulate a probabilistic version of IPFM and invert it using Bayes' rule to obtain another method to generate plausible time series of autonomic outflow from observed heart beats. However, here we focused on the GMC model for three reasons: i) the IPFM model has discontinuous jumps, which pose problems for Bayesian inference; ii) it cannot properly model ectopic beats; and iii) the GMC model supports a computationally efficient Gibbs sampler that allows for faster inference. Nonetheless, developing a Bayesian inversion of IPFM is a interesting avenue for future work.

\subsection{Code availability}

Code to run the Bayesian estimation of heart rate dynamics based on sequences of inter-beat intervals can be found in the repository \href{https://github.com/ferosas/BayesianAtHeart}{github.com/ferosas/BayesianAtHeart}.

\section*{Acknowledgements}

\noindent
We thank Daniel Bor, Sarah Garfinkel, and Andreas Roepstorff for inspiring conversations and helpful feedback. F.R. was supported by the Fellowship Programme of the Institute of Cultural and Creative Industries of the University of Kent. A.I.L. is supported by the Molson Neuro-Engineering Fellowship.

\newpage
\appendix

\section{Statistical results}
\label{app:stats}

\begin{table}[ht!]
\small
\begin{center}
\begin{tabular}{r c c c}
\multicolumn{4}{c}{}\\
\multicolumn{4}{c}{\textbf{\underline{Average heart rate}}}\\
\multicolumn{4}{c}{}\\
\specialrule{.13em}{.0em}{.15em} 
& Estimate & SE & p-value \\
\specialrule{.05em}{.08em}{.07em}
\textit{Frequentist}  &  & & \\    
\textbf{Self stand up}  &  10.61  &  0.93 &  $<$ 2e-16 *** \\
\textbf{Slow tilt}  &  9.27   &  0.88 &  $<$ 2e-16 *** \\
\textbf{Fast tilt}  &  10.14  &  0.89 &  $<$ 2e-16 *** \\
\specialrule{.03em}{.15em}{.0em} 
\textit{Bayes $\theta=0.01$}  &  & & \\
\textbf{Self stand up}  &  10.63  &   0.93 &  $<$ 2e-16 *** \\
\textbf{Slow tilt}  &  9.30   &   0.88 &  $<$ 2e-16 *** \\ 
\textbf{Fast tilt}  &  10.14  &   0.89 &  $<$ 2e-16 *** \\
\specialrule{.03em}{.15em}{.0em} 
\textit{Bayes $\theta=1$}  &  & & \\
\textbf{Self stand up}  &  10.50  &  0.93  & $<$ 2e-16 *** \\
\textbf{Slow tilt}  &  9.23   &  0.88  & $<$ 2e-16 *** \\ 
\textbf{Fast tilt}  &  9.98   &  0.89  & $<$ 2e-16 *** \\
\specialrule{.13em}{.15em}{.0em} 
\multicolumn{4}{c}{}\\
\multicolumn{4}{c}{}\\
\multicolumn{4}{c}{\textbf{\underline{Low frequency HRV}}}\\
\multicolumn{4}{c}{}\\
\specialrule{.13em}{.0em}{.15em}
& Estimate & SE & p-value \\
\specialrule{.05em}{.08em}{.07em}
\textit{Frequentist}  &  & & \\
\textbf{Self stand up}  &  -0.90$\times 10^{-4}$  & 1.65$\times 10^{-4}$  & 0.589   \\
\textbf{Slow tilt}  &  -4.97$\times 10^{-4}$  & 1.55$\times 10^{-4}$  & 0.00188 **\\
\textbf{Fast tilt}  &  -2.98$\times 10^{-4}$  & 1.58$\times 10^{-4}$  & 0.06238 . \\
\specialrule{.03em}{.15em}{.0em} 
\textit{Bayes $\theta=0.01$}  &  & & \\
\textbf{Self stand up}  &   0.36$\times 10^{-4}$ & 8.53$\times 10^{-5}$  & 0.67212   \\
\textbf{Slow tilt}  &  -2.56$\times 10^{-4}$ & 8.01$\times 10^{-5}$  & 0.00186 **\\
\textbf{Fast tilt}  &  -1.14$\times 10^{-4}$ & 8.14$\times 10^{-5}$  & 0.16308   \\
\specialrule{.03em}{.15em}{.0em} 
\textit{Bayes $\theta=1$}  &  & & \\
\textbf{Self stand up}  &  2.29$\times 10^{-5}$  & 5.92$\times 10^{-6}$  & 0.000196 ***\\
\textbf{Slow tilt}  &  -0.69$\times 10^{-5}$ & 5.56$\times 10^{-6}$  & 0.218629    \\
\textbf{Fast tilt}  &  0.13$\times 10^{-5}$  & 5.65$\times 10^{-6}$  & 0.821652  \\
\specialrule{.13em}{.15em}{.0em} 
\multicolumn{4}{c}{}\\
\multicolumn{4}{c}{}\\
\multicolumn{4}{c}{\textbf{\underline{High frequency HRV}}}\\
\multicolumn{4}{c}{}\\
\specialrule{.13em}{.0em}{.15em}
& Estimate & SE & p-value \\
\specialrule{.05em}{.08em}{.07em}
\textit{Frequentist}  &  & & \\
\textbf{Self stand up}  &  -8.59$\times 10^{-4}$  & 2.84$\times 10^{-4}$ &  0.00313 **\\
\textbf{Slow tilt}  &  -7.08$\times 10^{-4}$  & 2.67$\times 10^{-4}$ &  0.00922 **\\
\textbf{Fast tilt}  &  -7.99$\times 10^{-4}$  & 2.71$\times 10^{-4}$ &  0.00399 **\\
\specialrule{.03em}{.15em}{.0em} 
\textit{Bayes $\theta=0.01$}  &  & & \\
\textbf{Self stand up}  &  -5.35$\times 10^{-5}$  & 1.69$\times 10^{-5}$  & 0.002061 ** \\
\textbf{Slow tilt}  &  -4.21$\times 10^{-5}$  & 1.59$\times 10^{-5}$  & 0.009295 ** \\
\textbf{Fast tilt}  &  -5.92$\times 10^{-5}$  & 1.61$\times 10^{-5}$  & 0.000396 ***\\
\specialrule{.03em}{.15em}{.0em} 
\textit{Bayes $\theta=1$}  &  & & \\
\textbf{Self stand up}  &  -2.35$\times 10^{-7}$  & 9.07$\times 10^{-8}$  &    0.522\\
\textbf{Slow tilt}  &  -2.08$\times 10^{-7}$  & 8.52$\times 10^{-8}$  &    0.532\\
\textbf{Fast tilt}  &  -2.78$\times 10^{-7}$  & 8.66$\times 10^{-8}$  &    0.490\\
\specialrule{.13em}{.15em}{.0em} 
\end{tabular}
\end{center}
\end{table}

\begin{table}[ht!]
\small
\begin{center}
\label{tab:sym}
\begin{tabular}{r c c c}
\multicolumn{4}{c}{}\\
\multicolumn{4}{c}{\textbf{\underline{Permutation entropy}}}\\
\multicolumn{4}{c}{}\\
\specialrule{.13em}{.0em}{.15em} 
& Estimate & SE & p-value \\
\specialrule{.05em}{.08em}{.07em} 
\textit{Frequentist}  &  & & \\
\textbf{Self stand up}  & -0.009443 &  0.007571 & 0.215\\
\textbf{Slow tilt}  &  0.010748 &  0.007114 & 0.134\\  
\textbf{Fast tilt}  & -0.009659 &  0.007232 & 0.185\\  
\specialrule{.03em}{.15em}{.0em} 
\textit{Bayes, $\theta=0.01$}  &  & & \\
\textbf{Self stand up}  &  -0.005489 &  0.007582 & 0.471\\
\textbf{Slow tilt}  &   0.007487 &  0.007125 & 0.296\\ 
\textbf{Fast tilt}  &  -0.006723 &  0.007247 & 0.356\\ 
\specialrule{.03em}{.15em}{.0em} 
\textit{Bayes, $\theta=1$}  &  & & \\
\textbf{Self stand up}  &  -0.015936 &  0.001813 & 4.61e-14 ***\\
\textbf{Slow tilt}  &  -0.002576 &  0.001703 & 0.134\\
\textbf{Fast tilt}  &  -0.012141 &  0.001732 & 2.92e-10 ***\\
\specialrule{.13em}{.15em}{.0em} 
\multicolumn{4}{c}{}\\
\multicolumn{4}{c}{}\\
\multicolumn{4}{c}{\textbf{\underline{Hurst exponent}}}\\
\multicolumn{4}{c}{}\\
\specialrule{.13em}{.0em}{.15em} 
& Estimate & SE & p-value \\
\specialrule{.05em}{.08em}{.07em} 
\textit{Frequentist}  &  & & \\
\textbf{Self stand up}  &  0.078  &  0.043 & 0.0729 .  \\
\textbf{Slow tilt}  &  0.069  &  0.040 & 0.0897 .  \\
\textbf{Fast tilt}  &  0.045  &  0.041 & 0.2743    \\
\specialrule{.03em}{.15em}{.0em} 
\textit{Bayes, $\theta=0.01$}  &  & & \\
\textbf{Self stand up}  &  0.089  &  0.034 & 0.00965 ** \\
\textbf{Slow tilt}  &  0.053  &  0.032 & 0.10012  \\  
\textbf{Fast tilt}  &  0.043  &  0.032 & 0.18427\\
\specialrule{.03em}{.15em}{.0em} 
\textit{Bayes, $\theta=1$}  &  & & \\
\textbf{Self stand up}  &  0.188  &  0.017 & $<$ 2e-16 *** \\
\textbf{Slow tilt}  &  0.091  &  0.016 & 1.04e-07 *** \\
\textbf{Fast tilt}  &  0.159  &  0.016 & $<$ 2e-16 *** \\
\specialrule{.13em}{.15em}{.0em} 
\end{tabular}
\end{center}
\end{table}

\FloatBarrier

\end{document}